\documentclass{iaus}
\usepackage{graphicx}
\usepackage{url}

\newcommand{\DASCH}{{\it DASCH}~}
\def\Msun{$M_\odot$}

\title[\DASCH\/for Historical Time Domain Astronomy] 
{Opening the 100-Year Window for Time Domain Astronomy}

\author[J. Grindlay, S. Tang, E. Los \& M. Servillat]   
{Jonathan Grindlay, Sumin Tang, Edward Los, and Mathieu Servillat}

\affiliation{Harvard Observatory \& Center for Astrophysics, Cambridge, MA 02138, USA \\email: {\tt
  jgrindlay@cfa.harvard.edu}} 

\pubyear{2011}
\volume{285}  
\pagerange{1--6}
\jname{New Horizons in Time Domain Astronomy}
\editors{R.E.M. Griffin, R.J. Hanisch \& R. Seaman, eds.}

\begin{document}

\maketitle

\begin{abstract}
The large-scale surveys such as PTF, CRTS and Pan-STARRS-1 that have emerged
within the past 5 years or so employ digital databases and modern analysis
tools to accentuate research into Time Domain Astronomy (TDA).  Preparations
are underway for LSST which, in another 6 years, will usher in the second
decade of modern TDA. By that time the Digital Access to a Sky Century $@$
Harvard (\DASCH) project will have made available to the community the full sky
Historical TDA database and digitized images for a century (1890--1990) of
coverage. We describe the current \DASCH\/development and some initial results,
and outline plans for the ``production scanning'' phase and data distribution
which is to begin in 2012. That will open a 100-year window into temporal
astrophysics, revealing rare transients and (especially) astrophysical
phenomena that vary on time-scales of a decade.  It will also provide context
and archival comparisons for the deeper modern surveys.
  


\end{abstract}

\firstsection 
\section{Introduction}

As was clear from the 2010 US {\it Astronomy and Astrophysics Decadal Survey},
which ranked LSST as the highest-priority project for ground-based astronomy,
Time Domain Astronomy (TDA) has emerged as a key field of current astronomy and
astrophysics. The temporal domain offers new routes to astrophysical
understanding of extreme phases of stellar and galaxy evolution through studies
of nov\ae, supernov\ae, gamma-ray bursts and AGN, to list only a few.
Wide-field and/or temporal surveys of long duration can discover and study
binaries in various phases of evolution, some with exotic stellar components
such as black holes, or add to the increasingly rich harvest of exoplanets.
But while these extreme events or phenomena have short time-scales (hours to
months), astronomical time-scales are predominantly long, often measured in
millenia or even 10$^{3-6}$ times longer still.  Time-scales appreciably longer
than the duration of a given survey require the assimilation of large samples
of objects at different evolutionary (or binary) phases in order to piece
together temporal histories.  Now that we have the possibility to digitize
archived plate collections of large and (nearly) continuous duration and
totalling many images, it is possible to initiate and conduct studies in the
emerging field of {\it Historical TDA} for very large samples of objects on
time-scales that are at least an order of magnitude longer than was possible
before. That was the prime motivation for the
\DASCH\/project (Digital Access to a Sky Century $@$ Harvard).

\section{Development of \DASCH\/and Current Status}

The Harvard College Observatory plate collection is the world's largest,
containing some 450,000 direct plates.  As described elsewhere in these
Proceedings---see \pageref{Grindlay & Griffin}, it has approximately uniform
full-sky coverage from 1890 to 1990. The \DASCH\/project
(http://hea-www.harvard.edu/DASCH/) was initiated to digitize all the plates
and make their digital images ($\sim$400 Tb) and reduced photometric data
available on line (\cite[Grindlay et al.~2009]{Grindlay09}; Grindlay et al., in
preparation).  \DASCH\/incorporates the world's fastest and most
astrometrically precise plate scanner (\cite[Simcoe et al.~2006]{Simcoe06}) and
a powerful astrometric and photometric reduction pipeline (\cite[Laycock et
al.~2010]{Laycock10}). Over the 5-year development phase of the scanner and
reduction software, both the astrometry (\cite[Servillat et
al.~2011]{Servillat11}) and the photometry (Tang et al., in preparation) were
further optimized while scanning about 19,500 plates from five fields selected
for having calibrated sequences and a range of stellar densities.  A
semi-automated plate-cleaning machine is in the final stages of development; it
will clean the glass back-side of each plate more quickly than the 80 seconds
that it takes for an operating sequence of loading--scanning--unloading a
standard-size plate of 30\,cm $\times$ 25\,cm (they are scanned two at a time).
Full ``production scanning'' and processing of some 400 standard plates per day
can commence later this year when two such cleaning machines are
ready. Scanning of the whole collection of direct plates, together with
associated photometric reduction and population of the MySQL database, will
require approximately 4 years.  For each resolved object the database contains
positions (J2000), magnitudes, and a list of processing flags.  Full processing
into the database for all 400 plates scanned each day can be accomplished
overnight.

\section{\DASCH\/vs. Current and Planned TDA Surveys}

\begin{figure}[t]
\begin{center}
 \includegraphics[width=4.4in]{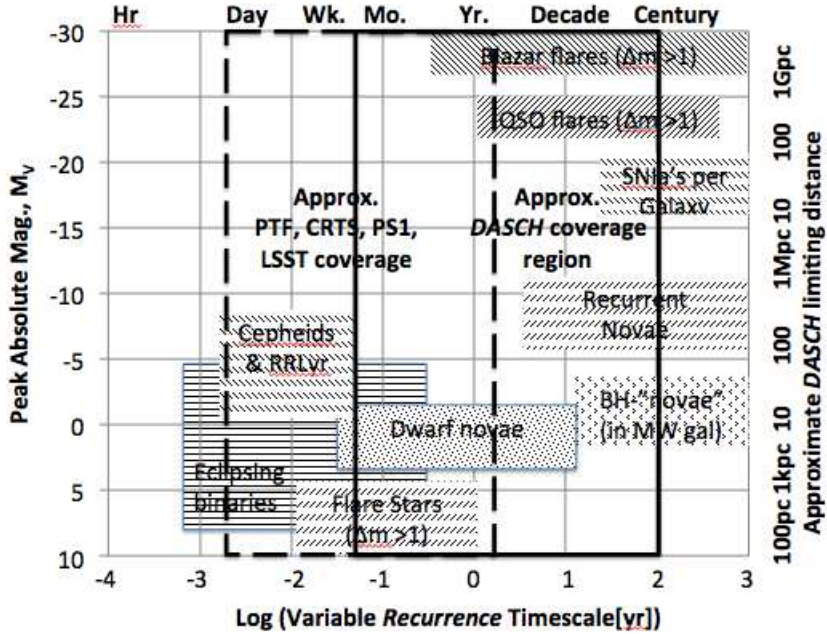} 

\caption{Representative classes of variables and transients vs. their
recurrence time that can be measured for a complete sample with
\DASCH\.(right) vs. PTF, CRTS, Pan-STARRS-1 and LSST (dashed box, left) or jointly
(overlap region). The distance scale (right axis) is for a conservative
\DASCH\/limiting magnitude $B$ = 14, which corresponds to $V$ = 15 for typical
colours.  For PTF, CRTS and LSST, the distance-scale axis for any object is
increased by a factor of 10--30 owing to their deeper limiting magnitudes. Note
that shorter variability time-scales may be measured by all those surveys, but
the short time-scale limits shown for each object class or survey are
approximately those for the variability to {\it recur}.}

   \label{fig1}
\end{center}
\end{figure}

In about 2 years the northern high latitude ($|b| >$ 10$^{\circ}$) full sky
from \DASCH\/will be on-line and available for access and analysis.  The
100-year temporal coverage, compared with $< $10 years of coverage by PTF and
CRTS and the several epochs of SDSS, will enable new studies of long time-scale
phenomena. Several examples are shown in Fig.~\ref{fig1}, where variable
classes are plotted by their approximate ranges of absolute magnitude (M$_V$)
against {\it recurrence} time-scale. Recurrence time-scales are chosen instead
of variability time-scales because they signal better which objects can be
discovered and measured.  Recurrence (or occurrence) is unambiguous, whereas a
variability time-scale can only be measured if there is nearly complete
temporal coverage; in practice, that is only available for a continuous-viewing
space mission like Kepler.  The PTF and PS1 surveys have typical observing
cadences (for a given sky region) of at least 1 day, though more frequent
sampling is being achieved for limited fields with PTF.  LSST will be
comparable, whereas for \DASCH\/the average observation cadence is about 2
weeks, though can be as short as 30 minutes on plates with multiple (offset)
exposures of a given field.  That sets the minimum recurrence time-scale that
can be measured; the maximum is limited by the total duration of the
survey. All the surveys can cover the same M$_V$ range, but of course the
corresponding distance range differs by the apparent magnitude limit of each
survey. We plot on the right of Fig.~1 an approximate distance scale out to
which each class of variable or recurrently variable object can be
detected. For \DASCH\/we adopted an approximate limiting magnitude of m$_V$ =
15, which is appropriate for an assumed ($B-V$) = 1; it corresponds to the mean
\DASCH\/limiting photographic ($B$) magnitude m$_B$ $\sim$14 measured on the
19,500 plates scanned thus far.

The recurrence times for various classes are taken from the literature, and are
either uncertain by the ranges shown (e.g., the recurrence time-scale range for
black-hole ``nov\ae'' is based on just 3 objects) or are an estimated range
given by systematic effects (such as the metallicity and and red-shift of
SN\,Ia hosts).  Recurrence-time ranges ending at the right side of the plot
(e.g., for bright blazar flares) are simply lower limits.  The recurrence times
for large flares ($\Delta$m $>$1) from flare stars or QSOs are estimates based
on variability studies against luminosity.  The overall conclusion is simply
that by expanding TDA surveys to time-scales that are 1 or 2 orders of
magnitude longer than those reached by current or immediately projected modern
surveys, a range of fundamental classes of objects can be studied as
``individual'' objects in well-defined (complete) samples.  Those could include
measuring the SN\,Ia rate in the Virgo cluster, or the outburst recurrence
times for the full set of the 20 or so currently known black-hole transients in
the Galaxy.  For each, recurrence time-scales can be measured or limited in the
``local'' Universe and then tested for red-shift dependence using much more
distant samples of differing objects in the modern TDA surveys.

\section{Current \DASCH\/Processing} 

\subsection{Photometry}
Photometric analyses developed by \cite[Laycock et al.~2010]{Laycock10}, and
recently further improved by Tang et al.~(in preparation), yield rms
uncertainties of $\sim$0.10\,mag over the full range of century of data from
the 9 or so series of plates that contribute to a typical light curve, despite
differences in plate scale, image quality and any systematic effects. The basic
approach employs SExtractor as the object detection and isophotal photometry
engine for instrumental magnitude determination by using the now global Hubble
Guide Star catalogue (GSC2.3) for a large sample of calibrators present on
every plate, thereby allowing both global and local calibrations.  Calibration
curves are first derived in annular bins to account for vignetting, by fitting
instrumental magnitudes against GSC2.3 magnitudes ($B$\/) for an initial
photometric solution. That initial calibration is followed by local corrections
to remove spatially-dependent plate effects (usually in the emulsion) or
sky-related effects (atmospheric extinction and clouds).  The GSC2.3 catalogue
is not ideal since its photometric precision is only $\pm$ 0.2 mag and it is
predominantly in a single band (photographic $B$\/). Fortunately, the all-sky
APASS CCD survey ({\tt www.aavso.org/apass}) described by \pageref{Henden} has
Johnson $B$\/ and $V$\/ as well as Sloan $g'$\/, $r'$\/ and $i'$, and will
improve significantly both the precision and, particularly, the colour
corrections for \DASCH photometry. APASS is now partly available and should be
full-sky by 2013.

\section{Representative Early \DASCH\/Results}

The five fields scanned in the \DASCH\/development phase are centred on M44,
3C273, Baade's Window, Kepler Field and the LMC. For the first three, plates
were simply selected if they contained the object (or the centre of Baade's
Window) on the plate, and at least 1 cm from the plate edge, or approximately
interior to ``bin 9'', the outermost annular bin or the outermost 5 mm of the
plate where both astrometry and photometry can have large errors. The coverage
obtained from that mode of plate selection was increasingly incomplete with
radial distance from the target object. For both the Kepler Field and LMC,
which are each extended regions, a wider boundary of plate-centre coordinates
was applied when selecting plates to ensure more complete coverage of the full
object. Only the M44 and Kepler Fields have been analyzed in detail, though
exploratory results have been obtained for the other 3 fields and will be
reported soon: for 3C273 (Grindlay et al., in preparation) and for a classical
nova discovered in Baade's Window (Tang et al., in preparation). The M44 study
led to the discovery of long-term dimming in a population of K giants
(\cite[Tang et al.~2010]{Tang10}). As a follow-up, more slowly-variable K
giants were found in the Kepler Field and studied with the higher time-coverage
of Kepler data (Tang et al., in preparation). The variability study of the
Kepler Field also led to the discovery of a dust-accretion event in the binary
star KU Cyg (\cite[Tang et al.~2011]{Tang11}).
 
\begin{figure}[b]
\begin{center}
 \includegraphics[width=3.4in]{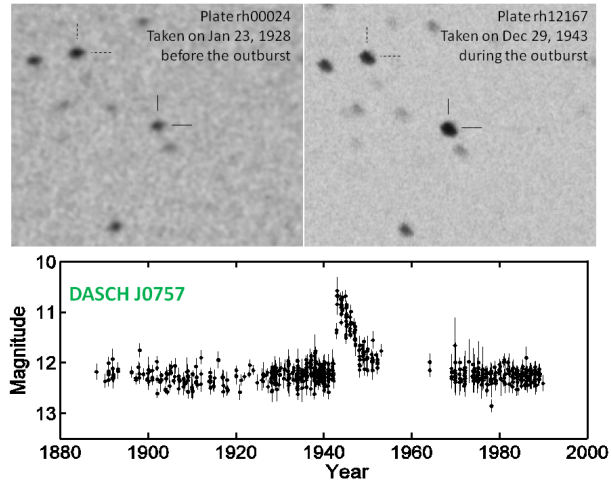} 
 \caption{Symbiotic nova images and light curve discovered by 
\cite[Tang et al.~2012]{Tang12}.  Upper panel: \DASCH\/images, showing
a quiescent phase (left) and an outburst (right).  The symbiotic nova
(marked) is in the centre; a comparison star (also marked) is to the
upper left.  Lower panel: Light curve derived from \DASCH\/scans.
The outburst might be due to H shell-burning on the white-dwarf
companion of the M0\,III giant in the system, although an
accretion-powered flare cannot be ruled out.}
 \label{fig2}
\end{center}
\end{figure}

A third, and perhaps most dramatic, example from \DASCH\/is a new type of
stellar variability, also discovered in the M44 field (Tang et al.~2012), and
is illustrated in Fig.~\ref{fig2}. Observations with the telescopes at the
Harvard-Smithsonian FLWO observatory in Arizona for spectroscopic
classification of \DASCH variables revealed that this star was an M0~III
giant. Comparison with the ASAS CCD photometric survey
(\cite[Pojmanski~2002]{Pojmanski02}) revealed it to be a semi-detached binary
with a 119.2-day period and an amplitude of 0.16 mag in the $V$ band.  As
described in detail by Tang et al.~(2012), the cool giant's companion is a
white dwarf, and we surmise that the remarkable flare in 1942 and its
subsequent 10-year decline was most probably due to nuclear H-shell burning.
The lack of emission lines from this symbiotic nova is new, but is consistent
with the behaviour of some other old nov\ae~and may also imply little or no
ejection of the envelope mass ($\sim$3 $\times$ 10$^{-5}$~\Msun) that is
required for ignition on what is probably a $\sim$0.6-\Msun\,white dwarf.

\begin{figure}[b]
\begin{center}
 \includegraphics[width=5.1in]{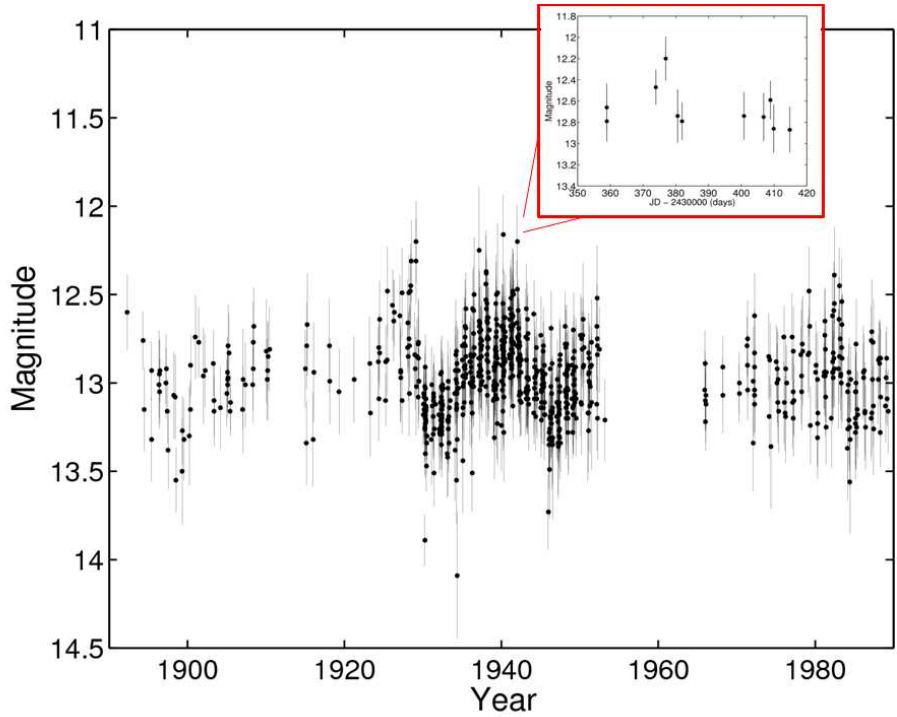} 
 \caption{Light curve of 3C273 from 786 points (of 1494 total) 
measured by \DASCH\/(Grindlay et al., in preparation). 
The 708 points not plotted are within 0.75 mag of their respective plate
limiting magnitude and may therefore have larger errors. One
example of a flare event over a 3-day time-scale is shown in the inset.
Note the absence of \DASCH\/measurements during the ``Menzel gap'' period 
from approximately 1955--1970, as discussed by Grindlay and Griffin 
\pageref{History} }
   \label{fig3}
\end{center}
\end{figure} 

We include a brief summary of the \DASCH\/light curve for the bright
quasar/blazar 3C273, which is discussed in detail by Grindlay et al.~(in
preparation). Its 100-year light curve (Fig.\,\ref{fig3}) demonstrates
the long-term variability of this luminous AGN.  The inset shows a ``flare'' 
in 1941 with characteristic rise and fall times of about 3 days (and there are
several other examples).  Time-scales of 1--30 days for brightening are 
evident in the overall light curve, and are comparable to the optical and
infra-red variability for 3C273 reported by \cite[Courvoisier 
et al.~(1988)]{Courvoisier88}, who found that the fastest variations were of
the order of 1 day but their observations  were too short to measure the 
longer-term variations detected with {\it DASCH}. The dominant power in the 
\DASCH\/variability spectrum is over longer time-scales like 0.5--2 years 
(such as the abrupt decline in flux from about 1927--1930), and provides
constraints on the size of the optical emission region.

\section{\DASCH\,Database} 
 
The \DASCH\/Pipeline (\cite[Los et al.~2010]{Los10}) and database software run
on a high-speed computer cluster and RAID disk system.  In production mode it
can process (overnight) the full Pipeline for the nominal 400 plates scanned in
a day, to populate a MySQL database with photometric values and errors for each
of the resolved stellar images; typically there are $\sim$50,000 on a standard
plate but there are more on ``A'' plates.  Light curves are generated very
rapidly for any object by extracting from the database the magnitudes thus
determined from all plates, or only those with magnitude measures meeting a set
of user-selected criteria.  Variability measures and tests of their validity
can then be derived readily. Additional variability analysis tools are being
developed, and will be made available when the full database becomes public.

The full \DASCH\/output database of $\sim$450,000 plate images and derived
magnitudes for each resolved object ($\sim$1Pb in total!) will be made
available for world access as it is completed incrementally.  The
present plans are to digitize the northern sky at Galactic latitudes $|b| >$
10$^{\circ}$ first, to allow comparisons with existing surveys such
as SDSS, PTF, Pan-STARRS-1 and the CRTS. That
stage could be completed by mid-2013. Given the difference in limiting
magnitudes (up to 14--18 for \DASCH\/against 20--22 for the modern surveys),
such comparisons will be mainly for context or extreme transients.  The
southern sky above/below the Galactic plane will come next, with expected
completion in early 2015, followed by the Galactic Plane by mid-2016---or still
well before LSST. The reason for doing the Galactic Plane and Bulge last is to
allow time to develop analysis of crowded-field photometry further by invoking
point-spread function (psf) and image subtraction techniques in order to
improve the present isophotal photometry used for SExtractor. Experiments have
recently been undertaken to optimise the use of PSFEx+SExtractor for magnitude-
and position-dependent fitting of the plate psf. \\ \\

\begin{acknowledgements}
We thank the \DASCH\/team and gratefully acknowledge support for \DASCH\/by the
HCO, the NSF (grants AST0407380 and AST0909073), and by the {\it Cornel and
Cynthia K. Sarosdy Fund for {\it DASCH}}.

\end{acknowledgements}

\end{document}